\documentclass[prb,preprint]{revtex4}
\usepackage{epstopdf}
\usepackage{rotating}
\usepackage{amsmath}
 
\begin{document}
 
\title{Lights Illuminate Surfaces Superluminally}
 
\author{Robert J. Nemiroff}
\email{nemiroff@mtu.edu}
\author{Qi Zhong}
\email{qizhong@mtu.edu}
\affiliation{Michigan Technological University, Department of Physics,
1400 Townsend Drive, Houghton, Michigan 49931}
\author{Elias Lilleskov}
\email{elillesk@macalester.edu}
\affiliation{Macalester College, Department of Physics \& Astronomy,
1600 Grand Ave., St. Paul, Minnesota, 55105}

\begin{abstract}
When a light bulb is turned on, light moves away from it at speed $c$, by definition. When light from this bulb illuminates a surface, however, this illumination front is not constrained to move at speed $c$. A simple proof is given that this illumination front always moves {\it faster} than $c$. Generalized, when any compact light source itself varies, this information spreads across all of the surfaces it illuminates at speeds faster than light. 
\end{abstract}
 
\maketitle

The Special Theory of Relativity does not allow objects with mass to accelerate past the speed of light.\cite{Einstein1905} Causal paradoxes may result from information transmitted away from its generation point at a speed faster than $c$, the speed of light.\citep{Benford1970} Nevertheless, it is well known that wave and illumination fronts may exceed the speed of light if they are not tied to mass or to transmitting locally produced information.\citep{Griffiths1994}

Two types of superluminal illumination front motion are possible. Perhaps the best known is created by a sweeping beam, commonly visualized by an illuminated spot from a laser beam moving along a wall. A conceptually similar situation is the shadow of an opaque object moving between a quiescent source and an illuminated wall. Sweeping beams can not only create spots that move superluminally, but generate unusual superluminal spot pair creation or annihilation events on the wall.\citep{Nemiroff2015}

A second type of superluminal illumination front occurs when any intrinsically variable source illuminates an object. For example, turn on a common light bulb and light will move away from this bulb at speed $c$: by definition the speed of light. However, illumination fronts from this bulb across surfaces are {\it not} constrained to speed $c$. As shown below, perhaps surprisingly, these surface illumination fronts are {\it always} superluminal.

A particularly simple proof of this starts by considering two points close together along the illumination front on a surface: a ``near point" relatively close to the bulb, and a ``far point" slightly further away. Examples are shown in Figure 1. Consider the points so close together that the surface is effectively flat between the points. The shortest path between the bulb and the far point is a straight line through the air -- the path that light takes to directly illuminate the far point. A longer path to the far point involves two lines -- a ``through the air" part from the bulb to the near point, and a ``along the surface" part from the near point to the far point. Of course, light moves through the air at speed $c$ to get from the bulb to both the near point and the far point. How, then, can the longer path end up illuminating the far point at the same time as the shorter path, given that the shorter path was completely traversed by light at speed c? The only way is if part of the longer path is traversed superluminally. Since the ``through the air" part of the longer path is traversed at speed $c$, the ``along the surface" part -- the part traversed by the illumination front -- must exceed $c$. Since these two points could be any points on the surface along the path of illumination, then all surface illumination fronts must move superluminally everywhere.

It is straightforward to compute the speed of the illumination front on an infinite planar wall. Assume that the bulb turns on instantaneously. Consider a line on this wall that passes as near as possible to the bulb. Assume that the light bulb occurs at a distance $s$ from $y = 0$, the closest point on this line, with distance along the line is given by $y$. Here time is tracked with the variable $t$ where $t = 0$ is here defined as the time the bulb turned on. The time that light from the bulb first strikes the line is at $t_1 = s / c$. From Figure 1 it is clear that $s^2 + y^2 = (c t)^2$ so that $y = \sqrt{ c^2 t^2 - s^2}$. Then the speed of the illumination front along the line is 
\begin{equation} \label{SpeedActualGeneral}
 v_y = {dy \over dt} = { c^2 t \over \sqrt{c^2 t^2 - s^2} } = {c^2 t \over y}  .
\end{equation} 
Note that when $y = 0$, $v_y$ is formally infinite! This means that when the light first illuminates the wall, the speed of the illumination front is not only superluminal -- it diverges. As time progresses the illumination front speed on the wall drops monotonically, only approaching -- but never reaching -- $c$ far from the initial impact point. Therefore, as indicated, the speed of the illumination front on the line is always superluminal.

It is also straightforward to use symmetry to generalize this argument from one dimension to two. The bulb will actually illuminate an outwardly moving circle on any illuminated wall. As this circle moves out, every possible surface orientation angle between the bulb and a surface will eventually occur. Since the illumination front speed is superluminal at all of them regardless of distance, then all object surfaces must show this effect, regardless of their distance, tilt, or orientation. Last, the illumination front does not have to refer to a light bulb turning on instantaneously -- the logic given here applies to the illumination fronts from any varying compact source.

Given the distance to the bulb and the time that the bulb turns on, tracking the illumination fronts this bulb creates on the surrounding walls would effectively map the distance and orientation of these walls -- a type of temporal tomography. This is because the distance between the observer and any location on an illumination front is uniquely determined by its observed angular distance from the bulb and its observed temporal delay from bulb onset. Therefore, although made difficult by the superluminal speeds, tracking illumination fronts has a direct application.

Stated generally and succiciently, when any compact light source itself varies, this information spreads across all of the surfaces it illuminates at speeds faster than light.

\begin{figure}[h]
\includegraphics[angle=90, width=18cm]{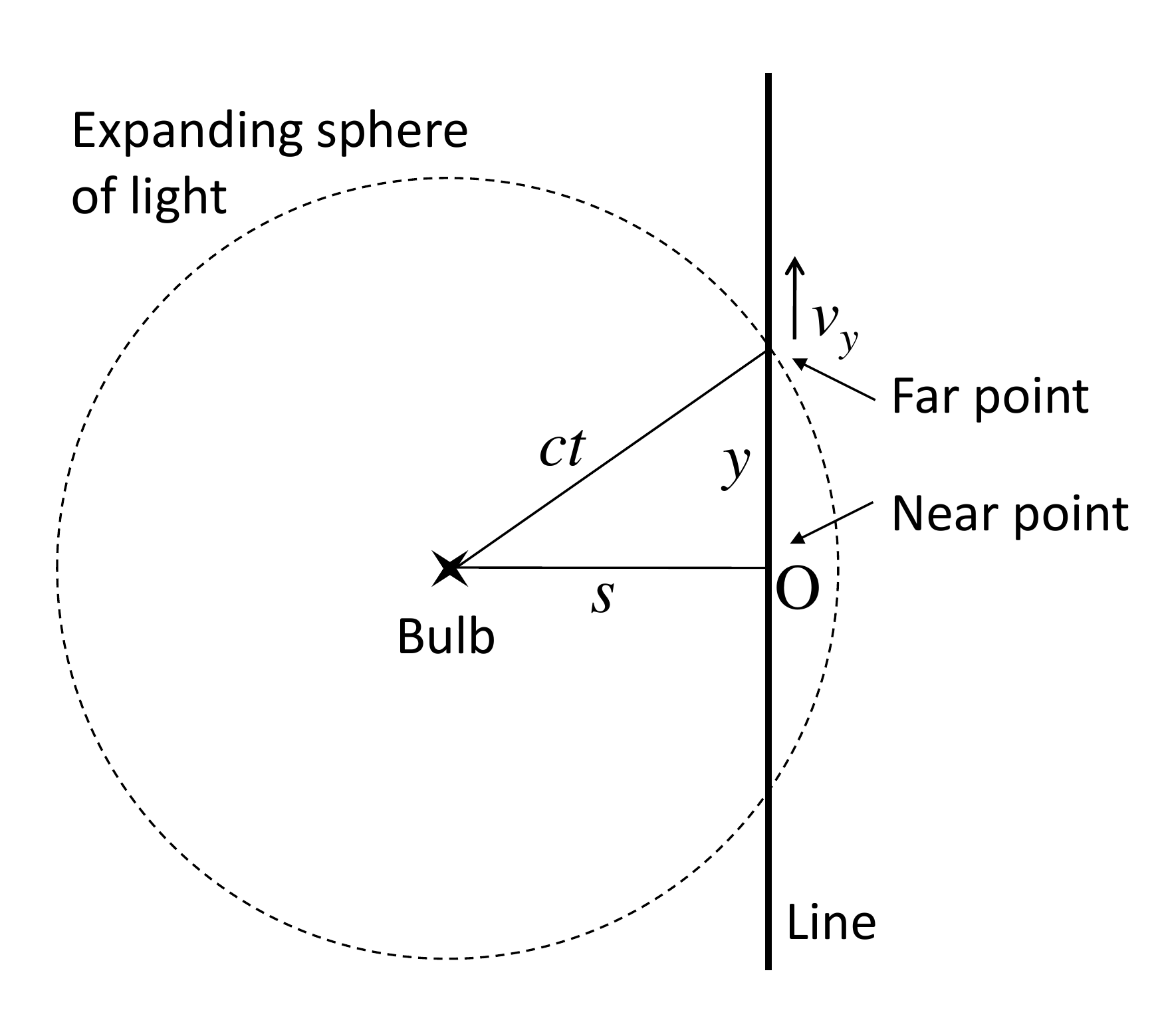}
\caption{The geometry of a illumination front expanding across a planar wall. The cross section for a line on the wall crossing closest to the illumination bulb is shown. When the initial expanding sphere of light, expanding at $c$, impacts the wall at $y = 0$, the initial speed of the illumination front, $v_y$ is formally infinite. As the illumination front moves along the wall, its speed drops monotonically toward $c$, only reaching $c$ asymptotically -- at infinite distance -- when the wall becomes parallel to light emitted from the bulb.}
\label{superfast}
\end{figure}


\begin{thebibliography}{25}
%
\bibitem[Einstein(1905)]{Einstein1905}
A. Einstein, ``Zur Elektrodynamik bewegter Korper", Annalen der Physik 322, 891--921 (1905).
%
\bibitem[Benford(1970)]{Benford1970}
See, for example, G.~A. Benford, D.~L. Book, and W.~A. Newcomb, ``The Tachyonic Antitelephone", Physical Review D 2, 263-265 (1970).
%
\bibitem[Griffiths(1994)]{Griffiths1994}
See, for example, D. J. Griffiths, ``Introduction to Quantum Mechanics" (First Edition), (1994).
%
\bibitem[Nemiroff(2015)]{Nemiroff2015} 
Nemiroff, R.~J.\ 2015, ``Superluminal Spot Pair Events in Astronomical Settings: Sweeping Beams", Pub. Astron. Soc. Australia, 32, e001 
%
\end{thebibliography}
\end{document}